\documentclass{ws-ijmpd}

\usepackage{amsmath}
\usepackage{latexsym}
\usepackage{amsfonts}
\usepackage{bm}
\usepackage{graphicx}
\usepackage{epsfig}
\usepackage{psfrag}

\newcommand{\be}{\begin{eqnarray}}
\newcommand{\ee}{\end{eqnarray}}

\newcommand{\mat}{\left ( \begin{array}{cc}}
\newcommand{\emat}{\end{array} \right )}
\newcommand{\vect}{\left ( \begin{array}{c}}
\newcommand{\evect}{\end{array} \right )}

\begin{document}


%
\catchline{}{}{}{}{}
%

\title{Equipartition of energy and the first law of thermodynamics at the apparent horizon}

\author{Fu-Wen Shu}

\address{College of Mathematics and Physics,Chongqing University of
Posts and Telecommunications, Chongqing, 400065, China\\
shufw@cqupt.edu.cn}

\author{Yungui Gong}

\address{College of Mathematics and Physics,Chongqing University of
Posts and Telecommunications, Chongqing, 400065, China\\
gongyg@cqupt.edu.cn}

\maketitle

\begin{abstract}
We apply the holographic principle and the equipartition law of
energy to the apparent horizon of a Friedmann-Robertson-Walker
universe and derive the Friedmann equation describing the dynamics
of the universe. We also show that the equipartition law of energy
can be interpreted as the first law of thermodynamics at the
apparent horizon. The consistency check shows that our derivation is
correct for $-1<w<-\frac13$, a value matches the recent cosmological
observations.
\end{abstract}


\section{Introduction}
The entropy of a black hole is 1/4 of the area of the event horizon
of the black hole measured in Planck units \cite{bekenstein}. This
area law of entropy was generalized to more general systems, and
later was taken as a general principle: the so called holographic
principle \cite{holo}. The strongest evidence of the holographic
principle was provided by the AdS/CFT correspondence which states
that all information about a gravitational system in a spatial
region is encoded in its boundary  \cite{maldacena}. On the other
hand, the thermodynamic laws of black holes and de-Sitter space-time
suggest a deep connection between gravitation and thermodynamics
\cite{bardeen}. By using the area law of entropy for all local
Rindler horizons, Einstein field equation was derived from the first
law of thermodynamics \cite{jacobson}.

Recently, the thermodynamic relation $S=E/2T$ \cite{padman} between
the entropy $S$, active gravitational energy $E$ and temperature $T$
was reinterpreted as the equipartition law of energy $E=N T/2$,
where $N$ is the number of bits on the horizon \cite{tp}. Combining
the equipartition law of energy and the holographic principle,
Einstein field equation was derived \cite{verlinde}. Furthermore,
gravity was explained as an entropic force caused by changes in the
information associated with the positions of material bodies
\cite{verlinde}, and this idea was extensively discussed \cite{fl}.

Motivated by Bekenstein's original thought experiment about black
holes, Verlinde considered a small piece of a spherical holographic
screen, and a particle with mass $m$ approaching it from the side,
with the change of entropy near the screen assumed to be
\cite{verlinde}
\begin{equation}
\label{deltas} \Delta S=2\pi \frac{mc}{\hbar}\Delta x.
\end{equation}
The effective force acting on the particle due to the change of
entropy is
\begin{equation}
\label{workeq} F\Delta x=T\Delta S.
\end{equation}
Because an observer in an accelerated frame has the Unruh
temperature
\begin{equation}
\label{unruht} T=\frac{1}{2\pi}\hbar a,
\end{equation}
so we get Newton's second law $F=ma$. Although it seems that we
derived Newton's second law for any force, actually the derivation
is satisfied for gravitational force only. Suppose the closed
holographic screen with radius $R$ can be divided into
$N=A/(c_1L_P^2)$ microscopic cells, where $L_P\equiv \sqrt{G\hbar}$
(we use units with $c=k_B=1$ throughout this paper) is the Planck
length and $c_1$ is a numerical factor whose value will be
determined later. If each cell has $c_2$ microscopic configurations,
then the entropy of the screen is \cite{tp}
\begin{equation}
\label{entropy} S=N\ln c_2=\frac{4\ln c_2}{c_1}\frac{A}{4L_P^2}.
\end{equation}
Once $4\ln c_2=c_1$ is chosen, the standard area law, $S=(A/4L_P^2)$
recovers. Recently it was shown that the entropy of a holographic screen
is $1/4$ of its area for static holographic screens if gravity is interpreted
as an entropic force \cite{kiselev}.
If one assumes that each cell of
area $c_1L_p^2$ contributes an energy $\frac12T$, according to the
equipartition law, we get the total energy
\begin{equation}
\label{energy} \mathcal{E}=\frac12 N T=\frac12 \frac{A}{c_1L_P^2}
T=M.
\end{equation}
so the Unruh temperature reads
\begin{equation}
\label{unruht1} T=\frac{2Mc_1 L_p^2}{4\pi R^2}=\frac{1}{2\pi}\hbar
a,
\end{equation}
and the acceleration is
\begin{equation}
\label{newtong} a=\frac{c_1 GM}{R^2}.
\end{equation}
If $c_1=1$, then we get the Newton's law of gravitation. So we
choose $c_1=1$.

By generalizing this approach to the relativistic case, Einstein
equation can be derived \cite{verlinde}. As we discussed above,
Einstein field equation can also be derived from the first law of
thermodynamics, and the Friedmann equation for several gravity
theories was derived from the first law of thermodynamics at the
apparent horizon (AH) \cite{cai,gong}, a natural question raised is
whether there is any connection between the first law of
thermodynamics and the equipartition law of energy. In this paper,
we address this problem by applying this approach to the apparent
horizon in cosmology. We first show that the first law of
thermodynamics at the apparent horizon is equivalent to the
equipartition law of energy at the apparent horizon, then we show
that Friedmann equation can be derived from the equipartition law of
energy and the holographic principle at the apparent horizon.
Finally, we make a consistency check of our derivation. It tells us
that our derivation is correct only when the state parameter
satisfies $-1<w<-\frac13$, a value matches the recent cosmological
observations. This constraint opens a door for applying Verlinde's
scheme to the study of dark energy.

\section{Derivation of the Friedmann equations}
\begin{figure}[h]\centering
\includegraphics[width=4.5in,height=2.5in]{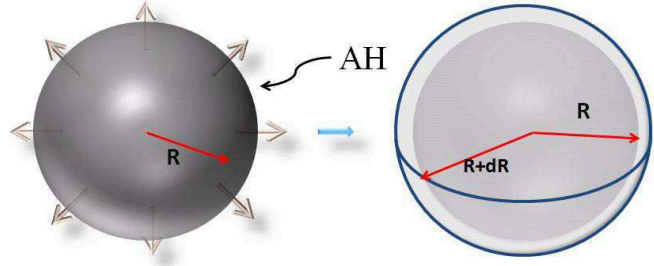}
\caption[]{Fluid flow out through the apparent horizon (AH) leads to
the increase of the radius of the AH from $R$ to $R+dR$.
}\label{fig1}
\end{figure}

Generally speaking, except the above two hypotheses, i.e., the
equipartition law and the holographic principle at the apparent
horizon, our derivation of the Friedmann equations needs one more
assumption: there is energy flux through the apparent horizon. The
basic idea is that by assuming an energy flux through the apparent
horizon with area $A$, the energy $\mathcal{E}$ which is enclosed by
the apparent horizon increases with the time. According to the
equipartition law, this leads to the changes of temperature and
number of bits of the screen, explicitly,\footnote{Note that this
does not mean that we are using two holographic screens with
different temperatures. Instead, we consider the same screen at an
infinitesimal later time with number of bits $(N+\Delta N)$ and
temperature $(T_A+\Delta T_A)$. Eq. (\ref{equiplaw}) is the
first-order approximation, i.e.,
$$
\Delta\mathcal{E}=\frac12(N+\Delta N)(T_A+\Delta
T_A)-\frac12NT_A=\frac{1}{2}N\Delta T_A+\frac{1}{2}T_A \Delta
N+higher\ orders.
$$}
\begin{equation}
\Delta \mathcal{E}=\frac12\Delta (NT)=\frac12T_A\Delta N+\frac12N
\Delta T_A,\label{equiplaw}
\end{equation}
where $T_A$ and $N$, respectively, represent the temperature and the
number of bits of the holographic screen.
By identifying $\Delta N$ with the changes of the area of the apparent
horizon $\Delta A$ through the holographic principle, we derive the
Friedmann equation. This is roughly sketched in figure (\ref{fig1}).
This counter-intuition picture can be well understood as following:
we take the apparent horizon as a holographic screen,
increase of the energy of the
region enclosed by a horizon inevitably leads to the increase of the
entropy of that region. Holographic principle tells us that this
corresponds to the increase of the horizon area. On the other hand,
the flux of fluid through the horizon implies our uncertainty about
the region outside the horizon increases. This is equivalent to the
increase of the entropy of the horizon, and again leads to the
increase of the horizon area.

To address our derivation more explicitly, let us focus on a
(3+1)-dimensional Friedmann-Robertson-Walker (FRW) universe with the
metric
\begin{equation}
 \label{2eq2}
 ds^2 = h_{ab}dx^adx^b +\tilde r^2 d\Omega_{2}^2,
 \end{equation}
 where $\tilde r = a(t) r$ and the
 2-dimensional metric $h_{ab} = {\rm diag} (-1, a^2/(1-kr^2))$ with $k=0,-1,1$
 corresponding to a flat, open and closed universe respectively. The apparent horizon, which
 is defined by the relation $h^{ab} \partial_a \tilde r
 \partial _b \tilde r=0$, turns out to be
\begin{equation}
 \label{2eq3}
 \tilde r_A = \frac{1}{\sqrt{H^2 +k/a^2}},
 \end{equation}
 where $H\equiv
 \dot a /a$ denotes the Hubble parameter. The
apparent horizon with area $A=4\pi \tilde{r}_A^2$ carries $N=4\pi
\tilde{r}_A^2/L_p^2$ bits of information. Suppose during the
infinitesimal time interval $dt$, the radius of the apparent horizon
evolves from $\tilde{r}_A$ to $\tilde{r}_A+d\tilde{r}_A$, then the
change of the area of the AH is $dA=8\pi\tilde{r}_A d\tilde{r}_A$.
So the number of bit is increased by
\begin{equation}
\label{dn} dN=\frac{8\pi\tilde{r}_A}{L_p^2}d\tilde{r}_A.
\end{equation}
On the other hand, the change of the Hawking
temperature\footnote{Although the definition of temperature for a
dynamic horizon like apparent horizon, as compared to a stationary
horizon, has not yet fully understood, it was shown
that there exists a Hawking
radiation with temperature $T_A=\hbar/2\pi \tilde{r}_A$ for locally
defined AH \cite{cai1}.} $T_A=\hbar/(2\pi \tilde{r}_A)$ is
\begin{equation}
\label{dT} dT_A=-\frac{\hbar}{2\pi\tilde{r}_A^2}d\tilde{r}_A.
\end{equation}
From Eq. (\ref{energy}), we get the changes of the total energy
(similar to Eq. (\ref{equiplaw}))
\begin{equation}
\label{dE} d\mathcal{E}=\frac{1}{2}NdT_A+\frac{1}{2}T_A
dN=\frac{d\tilde{r}_A}{G}.
\end{equation}
Note that the entropy of the apparent horizon is
$S_A=\pi\tilde{r}_A^2/L_P^2$, so $T_A dS_A=d\tilde{r}_A/G$. This
shows the equivalence between the equipartition law of energy and
the first law of thermodynamics, $d\mathcal{E}=T_A dS_A$. From the
definition of the apparent horizon (\ref{2eq3}), we get
\begin{equation}
\label{draeq}
d\tilde{r}_A=-H\tilde{r}_A^3\left(\dot{H}-\frac{k}{a^2}\right)dt.
\end{equation}

Now we discuss the energy flow through the apparent horizon within a
time interval $dt$. Because the energy-momentum tensor of the matter
in the universe is a perfect fluid, $T_{\mu\nu}=(\rho
+p)U_{\mu}U_{\nu} +p g_{\mu\nu}$, where $\rho$ and $p$ are  the
energy density and pressure, respectively, the energy flow out
through the horizon is
\begin{equation}
\label{dE1} -dE=d\mathcal{E}=4\pi \tilde{r}_A^2
T_{\mu\nu}k^{\mu}k^{\nu}dt=4\pi \tilde{r}_A^3(\rho+p)Hdt,
\end{equation}
where the (approximate) Killing vector or the (approximate)
generator of the horizon, the future directed ingoing null vector
field $k^{\mu}=(1,-Hr,0,0)$. Combining Eqs. (\ref{dE}),
(\ref{draeq}) and (\ref{dE1}), we get the following equation
 \begin{equation}
 \label{FE1}
 \dot H -\frac{k}{a^2}=-4\pi G (\rho +p).
 \end{equation}
Using the energy conservation equation
  \begin{equation}
  \dot \rho + 3 H(\rho +p) =0,
  \end{equation}
we get the Friedmann equation
  \begin{equation}
  H^2 +\frac{k}{a^2}= \frac{8\pi G}{3}\rho. \label{FE2}
  \end{equation}
In the above formulae, an integration constant, which can be
regarded as a cosmological constant, has been dropped out.

\section{Consistency check of the derivation}During our
derivation of the Friedmann equations, we have, as mentioned in the
last section, made an assumption that the energy is flowed out
through the apparent horizon. This is the only robust of our
derivation, and hence it deserves further investigation. To prove or
disprove this assumption one can make a consistency check of the
derivation\footnote{We would like to thank Professor Yong-Shi Wu for
his suggestion on this consistency check and for his valuable and
stimulated discussions.}. In other words, starting from the correct
Friedmann equations obtained by standard way (for example by solving
the Einstein' field equations), we consider if it is possible to
have an energy flux out through the apparent horizon.

For simplicity, we only consider $k=0$, cases with $k\neq 0$ can be
done in the same way. Then from the definition of $\tilde{r}_A$ we
obtain the expansion rate of the apparent horizon
\begin{equation}
\label{vofra1}\dot{\tilde{r}}_A=-\frac{\dot{H}}{H^2}.
\end{equation}
After combining the Friedmann equations (\ref{FE1}) and (\ref{FE2}),
this can be written as
\begin{equation}\label{vofra}
\dot{\tilde{r}}_A=\frac32(1+w),
\end{equation}
where we have introduced a parameter $w$ which is associated with
the equation of state $P=w\rho.$

To examine if the fluids will flow out through the horizon, we only
need to consider the fluid very close to the horizon. Recalling to
the definition of the apparent horizon which can be defined as the
boundary above which the fluid has velocities more than 1---the
velocity of light, we then obtain the velocity of the fluid (inside
the horizon) very close to the horizon (denoted by
$\dot{\tilde{r}}_f$)
\begin{equation}
\dot{\tilde{r}}_f=1-\varepsilon,
\end{equation}
where $0<\varepsilon\ll 1$. Comparing $\dot{\tilde{r}}_A$ with
$\dot{\tilde{r}}_f$ we are able to learn if the fluid flows out
through the horizon, i.e.,
\begin{equation}\label{energyflow}
\dot{\tilde{r}}_A-\dot{\tilde{r}}_f=\frac12+\frac32w+\varepsilon\simeq
\frac12+\frac32w.
\end{equation}
According to our derivation of Friedmann equations, two conditions
should be fulfilled: the flux of energy flows out through the
apparent horizon and the area of the horizon increases. This is
equivalent to require that $\dot{\tilde{r}}_A-\dot{\tilde{r}}_f<0$
and $\dot{\tilde{r}}_A>0$. From Eqs. (\ref{vofra}) and
(\ref{energyflow}) this can be satisfied when $-1<w<-\frac13$.
Therefore, our derivation of the Friedmann equations is
self-consistent if we impose a constraint on the state parameter by
$-1<w<-\frac13$, a value matches recent cosmological
observations\footnote{Actually, this constraint simply implies that
the universe is accelerated, since from
(\ref{vofra1})-(\ref{energyflow}), we see that
$$\dot{\tilde{r}}_A-\dot{\tilde{r}}_f \simeq-\frac{\dot{H}}{H^2}-1=-\frac{a\ddot{a}}{\dot{a}^2}<0$$
implies that $\ddot{a}>0$.}.

\section{Conclusions}

Motivated by the work on the origin of gravity \cite{verlinde}, we
derived the Friedmann equation from the equipartition law of energy
and the holographic principle. We also show that the equipartition
law of energy can be interpreted as the first law of thermodynamics
at the apparent horizon. This suggests that the equipartition law of
energy plays a fundamental role. Although there are many unresolved
issues on Verlinde's proposal, it has, at least in some extent,
transmitted a message that it is possible for the gravity to have a
thermodynamic origin, and our universe in this context is driven by
the so-called entropic force.

The consistency check shows that the derivation in our way is not
always effective---it will be invalid if the state parameter is out
of the region $-1<w<-\frac13$. This defect does not affect the
correctness of our derivation, on the contrary, the ``defect'' may
provide us with a natural requirement of dark energy \cite{shu}.

It also should be noted that our derivation of Friedmann equations
is different from Jacobson's previous work\cite{jacobson} in that:
(i)Our derivation is not restricted to the local Rindler causal
horizon, and a uniformly accelerated frame is not needed;
(ii)Instead of using the first law of thermodynamics, we are using
the equipartition law of energy.

{\bf Note added.} After the completion of our paper, there are two
relative works \cite{cai1,tp1} on the derivation, with different
methods, of Friedmann equations appeared.

\section*{Acknowledgments}

One of the author (FW) would like to thank Prof. Y.-S. Wu for his
stimulating discussions. This work was partially supported by the
NNSF key project of China under grant No. 10935013, the National
Basic Research Program of China under grant No. 2010CB833004, the
SRF for ROCS under Grant No. [2009]134, and the Natural Science
Foundation Projects of CQ CSTC under grant No. 2009BA4050 and
2009BB4084.

\section{References}


\end{document}